# Study of transient plasma stream using Time-Integrated Spectroscopic Technique

A.Ahmed[1], S. Singha[1], S. Borthakur[1], N.K.Neog[1], T.K.Borthakur[1]*, J. Ghosh[2]

∗ Corresponding author
E-mail address: tridip@cppipr.res.in (T.K.Borthakur)

[1]Centre of Plasma Physics-Institute for Plasma Research, Sonapur, Kamrup, Assam, 782402, India

[2]Institute for Plasma Research, Bhat, Ahmedabad, Gujarat, 382428, India

## Abstract

Time integrated spectroscopic measurements are carried out to characterize transient plasma stream produced in a coaxial pulsed plasma accelerator. This method allows the estimation of different plasma parameters and its evolution with time. It also provides information on the existence of different excited states from the spectral emissions of plasma. Using Argon as the discharge medium, the electron density estimated from Stark broadened line profiles gives a peak value ~$5 \times 10^{21}$ m$^{-3}$ at a discharge voltage of 15 kV and the flow velocity of the plasma stream is measured to be ~$(22 \pm 5)$ Km using Doppler shift method. Assuming p-LTE model, the electron excitation temperature is found to be ~0.88 eV using Boltzmann plot method. A temporal evolution of the plasma stream and its characteristic variation is studied from a time of 50 μs – 300 μs in steps of 50 μs by adjusting delay time in the triggering. Analysis of different spectral lines shows the existence of some metastable states of Ar II having a long lifetime. The evolution of different Ar II transitions to metastable and non-metastable lower levels is observed for different time frame. The temporal evolution study shows a decrease in electron density from $1.96 \times 10^{21}$ m$^{-3}$ to $1.23 \times 10^{20}$ m$^{-3}$ at 300 μs after the initiation of plasma formation. A decrease in excitation temperature from 0.86 eV to 0.72 eV is observed till 250 μs and then again rises to 0.77 eV at 300 μs.

**1. Introduction:**

Pulsed Plasma Accelerators (PPA) are a source of transient plasma which provide a high density, high velocity plasma stream. The typical nature of the plasma stream makes it front runner for many applications like plasma matter interaction (PMI) of fusion interest [1,2], fueling to fusion device [3], in material processing [4,5], space propulsion [6], study of astrophysical jets [7] etc. For these applications, detailed characterization of the plasma stream is necessary and this is carried out by using many diagnostic tools like electrostatic probes and Optical Emission Spectroscopy (OES), Interferometry, Thomson scattering etc. It is generally seen that the insitu measurements by electrostatic probes are widely used, but faces certain limitations as it obstructs the plasma.-On the other hand, OES is used by many researchers since it is a remote measurement technique where the scope of causing any perturbation to the plasma is completely ruled out. Morever, it is less complex and cost effective as compared to other optical methods which qualifies it to be a promising and reliable diagnostic in plasma studies [8–12]. Further, OES provides information about the constituents of plasma including the impurity particles from different elements present in the plasma. In addition to element identification, several works involving the use of OES for plasma parameter determination and study of temporal evolution is being reported [13,14]. This work presents time integrated spectroscopic measurements for different time frame of plasma life (time window) that are carried out to study the characteristics of the transient plasma stream produced in a pulsed plasma accelerator. In this method of OES measurement, the duration of emitted light collected from plasma can be changed adjusting the delay to the triggering of OES.

Therefore by adjusting the delay time to triggering with respect to initial bearkdown of the plasma, we can have collected time integrated emission spectrum of the plasma from the trigger time till the dying out time of the plasma.

Here, different plasma parameters such as plasma flow velocity, plasma density, and electron excitation temperature are estimated using this non-invasive spectroscopic technique. The initialization of the plasma and its evolution to till its quenching can be well understood by this study and hence dynamical behavior of plasma parameters can be well documented. The plasma stream observed in this way at different time window corresponding to different delay times has provided information on different spectral emissions. The temporal variation of density and excitation temperature of the plasma stream is obtained from the evolution study. This study also reveals the emission from different atomic particles of the working gas from different excitation and ionization levels. This information provides an insight to the equilibrium state of the transient plasma and its behavior with time evolution. A study of this kind is found to be limited only to few systems and lack an elaborate analysis of the subject in these type of transient plasma devices. This attempt is considered to be helpful in understanding the dynamics of a transient plasma stream having a comparatively longer lifetime than the conventional ones [1,15].

## 2. Experimental setup and diagnostics arrangement:

The pulsed plasma accelerator (PPA) is a two electrode system of cylindrical geometry. The inner electrode, which consists of 13 number of rods, is the cathode of the electrode system. The cathode is concentrically surrounded by an assembly of 17 numbers of rods to form the outer electrode. The outer electrode here is grounded and behaves as anode [16]. The whole experimental chamber is initially evacuated to a base pressure of $10^{-6}$ mbar with the help of a turbo-molecular pumping system. The PPA is powered by a 200 kJ pulsed power system (PPS), which delivers a damped sinusoidal current pulse for a time period of 1.0 ms with a peak discharge current of around 100 kA, while operating at 15 kV [17]. The transient high power pulse produced by the PPS is applied to the set of coaxial electrodes, which then energizes the plasma system. Argon is used as the working gas which is initially stored in the plenum or storage unit of an Electromagnetic Gas injection valve (GIV). A Trigger pulse generator [BNC 575] is used for synchronized triggering of the main discharge pulse of PPS and the gas injection from the GIV. Different delays can be set in the trigger generator which enables triggering of different systems at different times. A delay of ~ 4 ms is set between the triggering of the PPS and the GIV, so that the discharge occurs when a substantial amount of gas particles reaches the discharge initiation point in the annular region of the electrodes. On triggering the GIV, the gas is injected into the chamber through 24 nos. of nozzles present between the coaxial cathode and anode assemblies. The breakdown of the gas forms a plasma sheath in the annular region. The plasma sheath moves towards the open end of the electrode by the **J×B** force where **J** is the current density in the plasma sheath and **B** is the self generated magnetic field. The diffuse plasma sheath moving during the discharge pulse forms the plasma stream and while leaving the electrode the stream is compressed by the radial component of **J**×**B** force. The experimental studies in this work is carried out by collecting light from this region of compression zone at the exit of electrode assembly. This is performed by an optical system arrangement which provides an emission spectrum as described in next section.

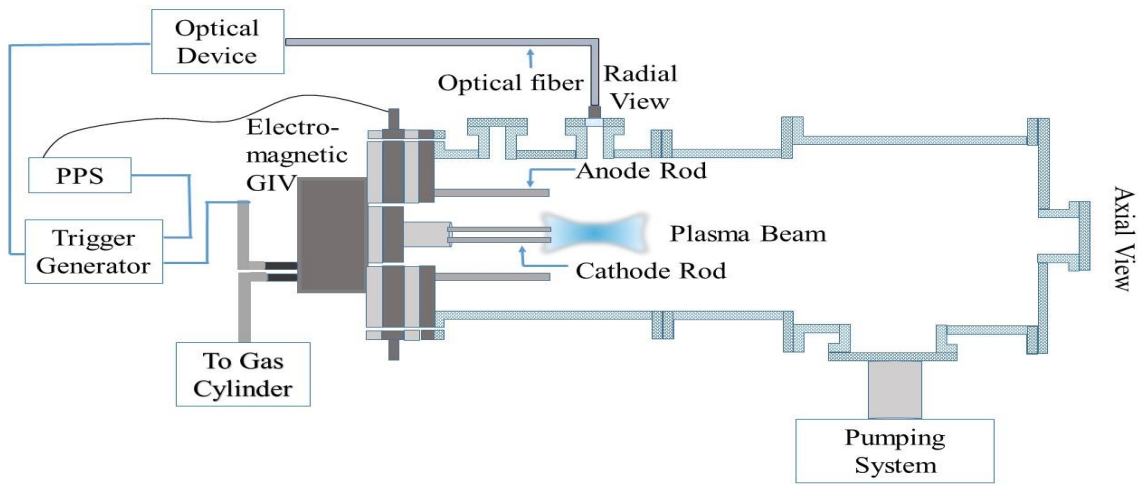

**Fig.1. Schematic of Pulsed Plasma Accelerator along with diagnostic arrangement**

The main part of the of the OES system is a spectrometer working in the visible range of light spectrum. The spectrometer [Avantes AvaSpec-Dual] used here is equipped with a grating of 1200 lines/mm and 10 µm slit and gives a resolution of 0.1 nm at a blazing wavelength of 500 nm. The sensitivity of the detector lies in the visible spectral wavelength range of 350 - 820 nm. The main objective of the work is the identification of the emitted lines from the plasma, the determination of the plasma parameters and the study of the behavior of plasma stream in different time window of its life span. An optical fiber is positioned and mounted on a side viewport of the chamber so that it collects light emitted in the radial direction from the plasma stream just at the exit region of electrodes end. The other end of the fiber is coupled to the spectrometer. The triggering of the spectrometer is carried out through the trigger pulse generator, which is synchronized wih the discharge pulse of PPS. The trigger pulse to the spectrometer is adjusted by varying the delay time with reference to discharge current pulse and this allows the spectrometer to capture the spectrum of the plasma stream in different time window. Although a time-integrated spectrum is obtained, spectral data of plasma can be collected for different period of plasma lifetime from the trigger time to the end of the plasma life. Prior to collecting spectra from plasma, the spectrometer is calibrated for wavelength with a standard calibration source of Hg-Ar. It is noted that there is no other broadening mechanism in the calibration source so the broadening of the spectral lines emitted from Hg-Ar atoms is purely due to the spectrometer. This instrumental broadening of the spectrometer is calculated from the FWHM ($\lambda_I$) of the Gaussian profile of the Hg-Ar spectral lines and is found to be 0.15 nm. After calibration, the emission spectra of plasma beam were collected at two different discharge voltages of 8 kV and 15 kV and their spectra were analyzed for deriving various plasma parameters such as plasma flow velocity, plasma density and electron excitation temperature. For our convenience, the experiments to collect spectral information for different time window are conducted at lower discharge voltage of 8 kV. The pressure of the plenum of GIV is set at 4 bar for whole experimental work.

## 3. Results and Discussions:

### 3.1 Emission spectrum for the entire plasma lifetime and its measurement

The line emission spectrum is obtained during the entire plasma lifetime as shown in fig. 2. Here, the Atomic Spectra Database of National Institute of Standards and Technology (NIST) is used for identification of spectral lines [18]. The emission spectrum obtained from the plasma in PPA shows mostly the lines from singly ionized argon and few other impurity atoms. The singly ionized argon lines are found to be intense and dominant in the spectrum. These lines are mostly located in the wavelength range of 400-500 nm. Afew low intense lines from neutral excited argon atoms are also observed. These low intense lines lie in the higher wavelength side of the spectrum. Moreover, two neutral lines (763.51 nm, 811.53 nm) are also observed (as shown in emission spectrum) while the intensity of other neutral lines are comparatively low. There are some impurity lines and they are identified to be of Fe (354.71 nm, 356.06 nm, 373.13 nm, 378.19 nm, 385.43 nm, 387.17 nm, 656.53 nm), C (657.80 nm), Mn (357.78 nm) as labeled in the spectrum shown in fig. 2. These impurity lines are expected in PPA. This is due to the dragging of the plasma sheath during its acceleration down the electrode channel and in the process it erodes electrodes surface. The insulator surface coming into the contact of plasma sheath also gets some surface erosion. An increase in impurity level is obtained in the plasma produced by higher discharge voltage. Thus, it can be infered that erosion of electrodes and insulator becomes much higher if we apply higher discharge voltage to produce the plasma stream.

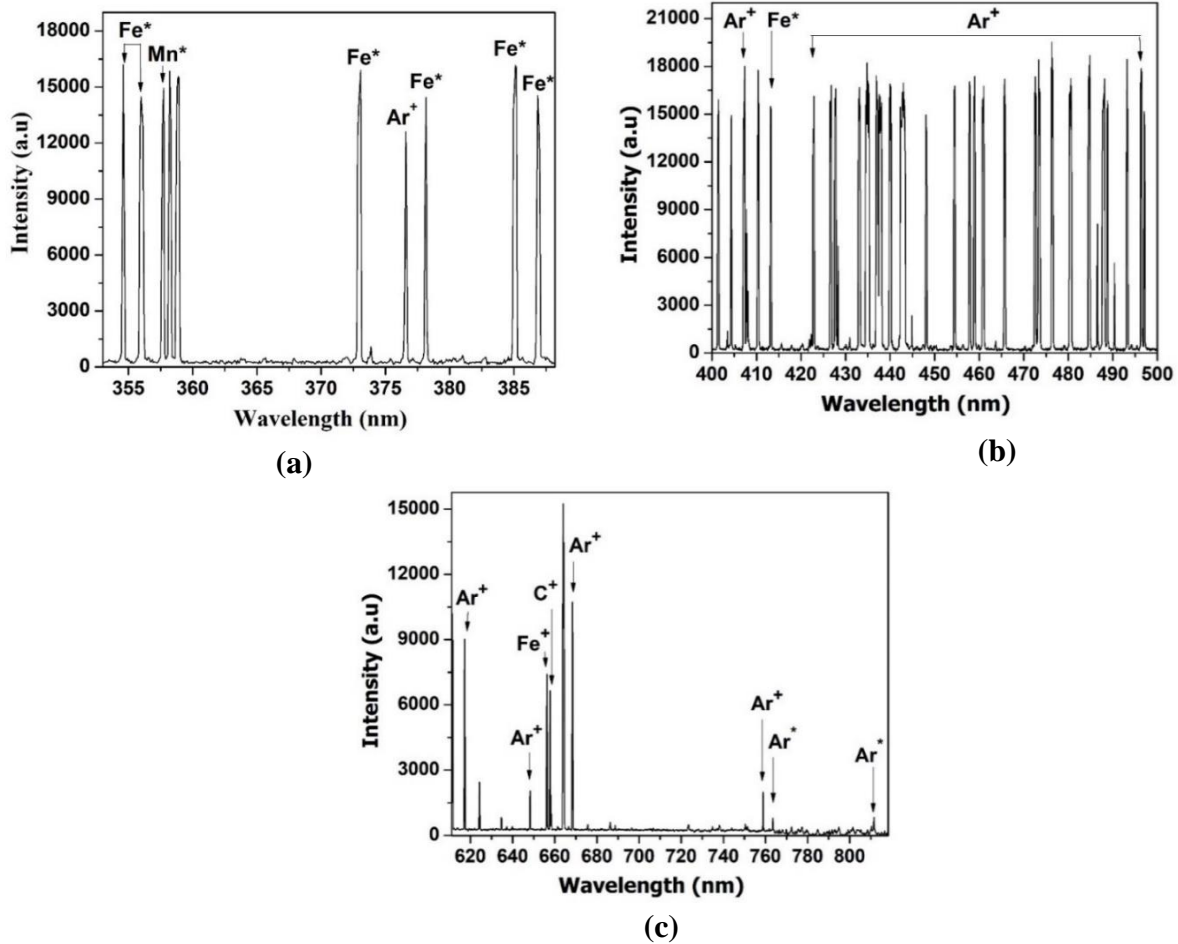

**Fig. 2 (a), (b), (c) showing the emission spectra from argon plasma with labeled argon ions, atoms, and impurity atoms**

Another important observation by optical emission spectroscopy is the broadening and shift that may occur in an emitting plasma and these measurements can give interesting information about electron density, electron temperature, electron excitation temperature and plasma flow velocity. Thus the broadening and shift of the spectral lines which are observed from our plasma stream also, has given important information about it. The width of the spectral lines occurs due to different broadening mechanisms like Doppler broadening, natural line broadening, pressure broadening, and Instrumental broadening. The Doppler broadening is caused due to thermal motion of the emitters whereas the natural line broadening arises as a result of uncertainty principle. The pressure broadening includes stark broadening caused by perturbations due to charged particles, Vander Waals' broadening occurring due to the interaction of emitter dipole with the dipole of the surrounding perturbing neutrals and resonance broadening due to dipole-dipole interaction of emitter atoms with the ground state atoms of the same element [19–21]. The Instrumental broadening arises due to finite resolution of the spectrometer and can be calculated as mentioned earlier. Besides broadening and shift, dip or flat top in the spectral lines is also observed which indicates presence of self-absorption. This occurs due to the absorption of radiation by the cold atoms present in the periphery of the central hot core of plasma stream. The self-absorbed lines are known as optically thick lines where the intensity of the emissions decreases and the broadening of the lines increases [22,23]. For measurement pupose, in this work, we have considered only the lines which don't have flat top or dip structures, which implies optically thin condition. The self absorbed lines are discarded as it results in overestimation of plasma parameters. For analysis, we fitted the spectral lines to voigt profile, which is a convolution of Lorentzian and Gaussian profile.

### 3.1.1 Plasma stream velocity

As mentioned earlier, the plasma sheath formed in the inter-electrode region of the pulsed plasma accelerator is accelerated due to Lorentz force. This results in a high flow velocity of plasma stream moving towards the outer end of the electrode assembly. For the estimation of this plasma flow velocity, the Doppler shift method of light emission is employed [24]. This method involves the shift in light frequency due to relative motion of the emitter particles or source of emission in the plasma. The emission spectrum is recorded for two different directions of observations. The light emission is first collected from the open end of the electrodes system in radial direction where line of sight is perpendicular to plasma flow direction. The measurements are then taken along the axial direction where the line of sight is parallel to the direction of motion of the plasma particles. As the plasma stream moves along the axial direction, there is movement of the emitting particles towards the detector which causes a blue Doppler shift in the wavelength of the line emission. The wavelength of line emission obtained in radial view is considered as the reference wavelength. During the axial view, the shift in wavelength observed with respect to the reference wavelength, is used to calculate the velocity of the emitting particles by using Doppler shift formula. The Doppler shift ($\Delta\lambda$) of spectral line is given by equation (1) as-

$$\frac{\Delta\lambda}{\lambda} = \frac{v}{c} \times \cos\theta \qquad (1)$$

Where '$\lambda$' is the reference wavelength in nm, '$v$' is the velocity of the emitting particles in m/s and '$\theta$' denotes the angle between line of sight and direction of motion of the emitting particles.

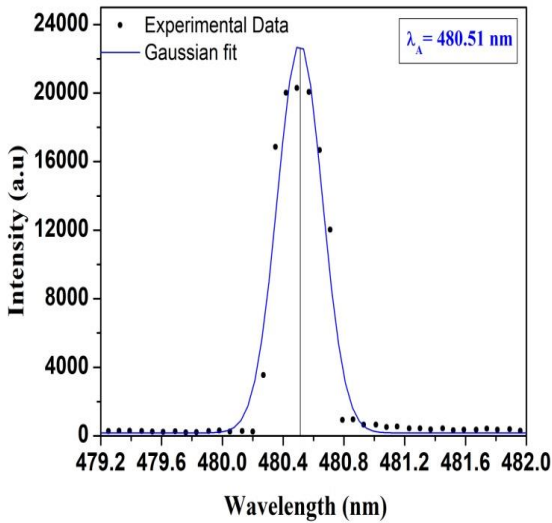

**Fig. 3(a) Spectral Line of Ar II in Radial Observation obtained at maximum operating voltage of 15 kV considered as reference wavelength**

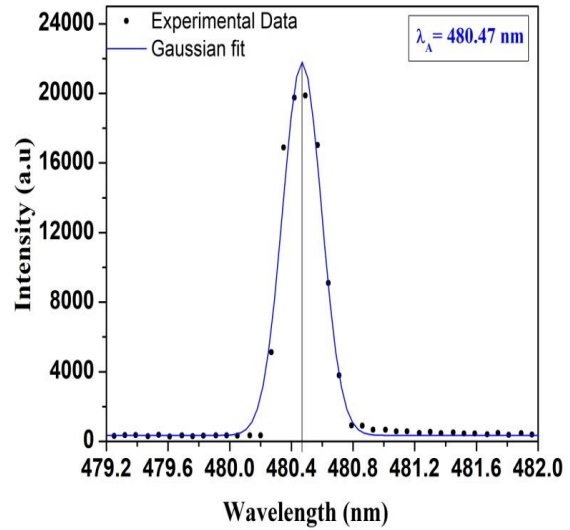

**Fig. 3(b) Spectral Line of Ar II in Axial Observation obtained at maximum operating voltage of 15 kV considered as shifted wavelength**

In the axial view, the line of sight is along the direction of motion of the emitting plasma particles and this gives a value of unity for the cosine term in eqn. (1). From the emission spectrum obtained in both directions, a prominent Ar II spectral line of wavelength 480.60 nm is identified using the NIST database.

Since simultaneous observation is not possible in our case, spectra of 3 shots (1 Shot refers to a single gas discharge for plasma formation in PPA) were analyzed for plasma stream velocity at same experimental condition of PPA. Using the relation (1), the average velocity of the plasma stream flow, obtained from 3 pair of consecutive shots taken in radial and axial directions, is estimated to be $22 \pm 5$ km/s. Similarly, five other spectral lines (434.80 nm, 473.59 nm, 484.78 nm, 487.98 nm, 500.93 nm) from the same spectra emitted from argon plasma are analyzed to find the velocity. The observation shows a variation of plasma flow velocities in the range 10-30 km/s. Similar variation of velocity in spectroscopic measurements were also reported by others [25]. It is worth mentioning that the plasma flow velocity (~20 Km/s) was theoretically estimated in this PPA and experimentally (~ $25 \pm 1.5$ Km/s) by using Double plate probe, which corroborates our estimation in this experimentation [16,26].

### 3.1.2 Plasma Density

In plasma, the local micro electric fields exist due to the presence of charged particles and it influences or interacts with the surrounding plasma species such as ions or neutral atoms. This electric fields perturb the degenerate energy levels of the surrounding atoms or ions. This electric field breaks or lifts the degeneracy of the energy levels and splits it into different levels of slightly different energies. This is known as the Stark effect which can be commonly observed in plasma of density >$10^{20}$ m$^{-3}$ [27,28]. The plasma stream of PPA is a high density plasma as estimated earlier by using Triple Langmuir probe which is of the order of $10^{21}$ m$^{-3}$ [29]. Due to the higher density of plasma, Stark effect is likely to be dominant than the other broadening mechanisms. The strong electric fields alter the emission process of the neutrals and ions as the atomic transitions taking place from different closely spaced energy levels result in broadening and shift of the spectral lines. Obradovic et al. measured the macroscopic electric field strength by the stark shift caused by macroscopic electric field in Magnetoplasma compressor (MPC) device [30].

Different models are put forward in literature for the analysis of observed spectral emissions. The measurement of electron density by Stark broadening is an indirect method which depends on the assumption of the model employed. The broadening and shift of the spectral emissions is due to the combined effect of electric fields from ions and electrons. As the high mobility of electrons result to fast collisions, therefore an electron impact approximation is used to study the effects on line broadening. On the other hand, due to slow-moving ions, a static electric field is assumed which however depends on the timescale of the electric field variation and the timescale of the emitter dipole interaction. For a larger field variation timescale, a quasi-steady model can be assumed. The broadening of spectral lines caused by impact of fast electrons is symmetric and follows a Lorentzian profile while those influenced by heavy ions have an asymmetric profile [31]. For hydrogen atom or for atoms in which the energy difference between perturbed energy levels is small in comparison to the interaction energy between emitter and surrounding perturbing plasma particles, the Stark effect is linear. For other non-hydrogenic atoms, the Stark effect is quadratic. The electron density, $N_e$ (cm$^{-3}$) can be calculated using the equation (2) which considers both electron contribution (first part of the equation) and ion contribution (second part of the equation) to the line broadening [32,33].

$$\Delta\lambda_L = 2\,\omega \left[\frac{N_e}{10^{16}}\right] + 3.5\,A \left[\frac{N_e}{10^{16}}\right]^{\frac{5}{4}} \left[1 - \frac{3}{4} N_D^{-\frac{1}{3}}\right] \omega \qquad (2)$$

Here, $\Delta\lambda_L$ is the Lorentzian FWHM of the spectral line in nm, $\omega$ is the electron impact parameter in nm, $A$ is the ion broadening parameter in nm, $N_D$ is the number of particles in the Debye sphere. It is seen that, $H_\alpha$ and $H_\beta$ of hydrogen are the most widely used lines for calculation of electron density as the highly degenerate states of the hydrogen atoms leads to a linear stark effect. In some experiments, hydrogen gas is mixed in lower proportion with the working gas to obtain the $H_\alpha$ and $H_\beta$ lines for analysis purposes. In our work, there is no hydrogen mixing with the working gas of argon and the spectral lines of argon are used for analysis. Some of the most degenerate levels of a particular emission are highly perturbed by the intense electric fields resulting in dip structures and hence cannot be used for analysis purpose. It is found that the symmetric singly ionized argon intense line of 454.50 nm [$3p^4$ ($^3P$) $4p - 3p^4$ ($^3P$) $4s$] is free from such dip structure and thus it is used to calculate the electron density from its stark broadening. The ion contribution to the broadening is considered to be negligible as no asymmetry is observed in the selected line profile .

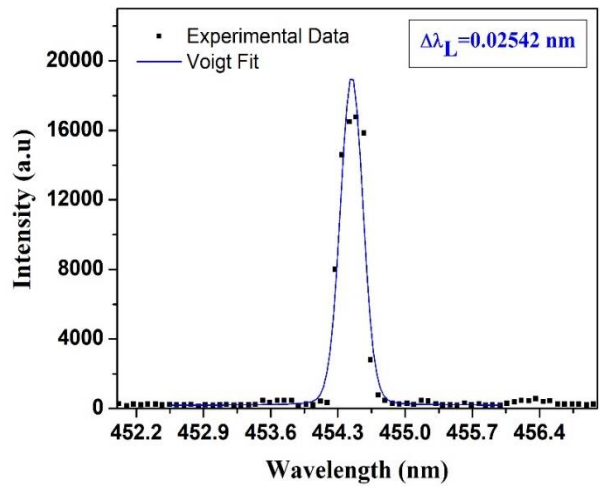

**Fig. 4. Voigt fitted profile of 454.50 nm Ar II spectral line obtained at 15 kV discharge voltage**

Hence for dominant electron contribution, the second part of eqn. (2) vanishes. The equation thus reduces to the form (3)

$$\Delta\lambda_L = 2\,\omega\left[\frac{N_e}{10^{16}}\right] \quad (3)$$

The FWHM of Lorentzian profile of the spectral line is obtained after de-convolution of its voigt profile as shown in fig. (4). Using this Lorentz width $\Delta\lambda_L$ (0.02542 nm) and value of $\omega$ [34], the plasma density is estimated from the equation (3). The estimated plasma density, which is averaged over for three consecutive shots, is found to be around $(5.29 \pm 0.38) \times 10^{21}$ m$^{-3}$. This value of plasma density is very close to the experimentally measured data using Triple Langmuir probe in this device [29]. Similar density has already been observed for plasma streams of other pulsed plasma devices like QSPA Kh-50, QSPA-M, MPC etc. [12,35].

### 3.1.3 Electron excitation temperature

The excitation temperature in plasma arises due to the transfer of kinetic energy of electrons during collisions with the ions or neutrals. It can be assumed here that the plasma is in partial local thermodynamic equilibrium (p-LTE) where the plasma cannot be defined by a single temperature and the electron temperature and excitation temperature are related in this case as $T_e > T_{exc}$ [36]. The electrons in plasma establish equilibrium between different upper-level populations of the higher excited ions. Under the assumption of p-LTE, the excited level populations follow a Boltzmann distribution and the emission intensity from these excited levels can be expressed as-

$$\frac{I_{ji}\lambda_{ji}}{A_{ji}g_j} = \frac{hcN}{4\pi U(T_{exc})}\exp\left[-\frac{E_j}{KT_{exc}}\right] \quad (4)$$

Where, $\lambda_{ji}$ (m) is the transition wavelength, $I_{ji}$ is the line intensity, $A_{ji}$ (s$^{-1}$) is the transition probability from $j$ to $i$ level, $g_j$ is the statistical weight of the upper energy level $j$, $E_j$ (eV) is the energy of the upper excited energy level $j$, $N$ is the population of the ground state, $U(T_{exc})$ is the partition function, $KT_{exc}$ gives the electron excitation temperature in eV.

Table 1: Spectroscopic data of 7 singly ionized Argon emission spectral lines obtained from NIST database [18]

| Wavelength $\lambda_{ji}$ (nm) | Transition Probability $A_{ji}(\times 10^7$ S$^{-1})$ | Statistical weight $g_k$ | Upper energy level $E_k$(eV) | Transition levels L.L Configuration, Term, J – U.L Configuration, Term, J | Observed Relative Intensity |
|---|---|---|---|---|---|
| 426.65 | 1.64 | 6 | 19.54 | 3p$^4$($^3$P)4s, $^4$P, 5/2 – 3p$^4$($^3$P)4p, $^4$D$^o$, 5/2 | 16818 |
| 454.50 | 4.71 | 4 | 19.86 | 3p$^4$($^3$P)4s, $^2$P, 3/2 – 3p$^4$($^3$P)4p, $^2$P$^o$, 3/2 | 16743 |
| 457.93 | 8.00 | 2 | 19.97 | 3p$^4$($^3$P)4s, $^2$P, 1/2 – 3p$^4$($^3$P)4p, $^2$S$^o$, 1/2 | 17051 |
| 460.95 | 7.89 | 8 | 21.14 | 3p$^4$($^1$D)4s, $^2$D, 5/2 – 3p$^4$($^1$D)4p, $^2$F$^o$, 7/2 | 16754 |
| 496.50 | 3.94 | 4 | 19.76 | 3p$^4$($^3$P)4s, $^2$P, 1/2 – 3p$^4$($^3$P)4p, $^2$D$^o$, 3/2 | 17834 |
| 500.93 | 1.51 | 6 | 19.22 | 3p$^4$($^3$P)4s, $^4$P, 3/2 – 3p$^4$($^3$P)4p, $^4$P$^o$, 5/2 | 19196 |
| 506.20 | 2.23 | 4 | 19.26 | 3p$^4$($^3$P)4s, $^4$P, 1/2 – 3p$^4$($^3$P)4p, $^4$P$^o$, 3/2 | 18806 |

Taking logarithm on both sides of equation (4), the new form is expressed in equation (5) -

$$\ln\left(\frac{\lambda_{ji} I_{ji}}{A_{ji} g_j}\right) = \left(-\frac{1}{KT_{exc}}\right)(E_j) + \text{Constant} \qquad (5)$$

Here, (-1/KT$_{exc}$) is the slope of the Boltzmann plot and its inverse gives the electron excitation temperature. The spectral lines corresponding to different transitions of ionized argon used for the estimation of excitation temperature are listed in Table. 1 along with other parameters.

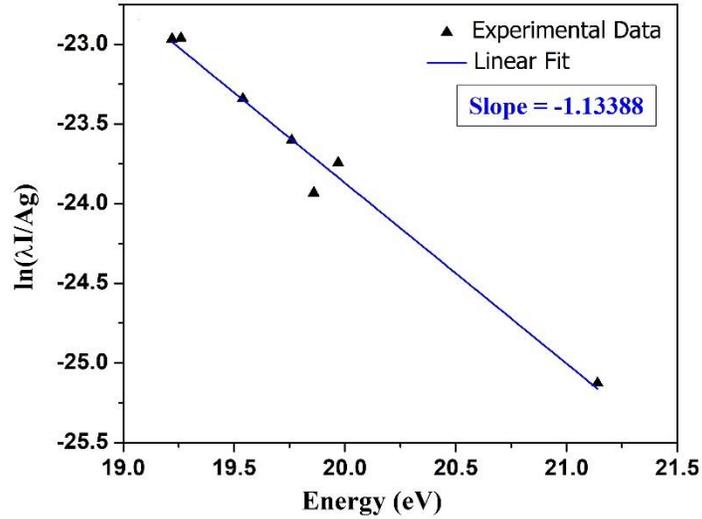

**Fig. 4. Boltzmann Plot obtained using data points listed in Table.1 for different Ar II spectral lines**

The slope of the plot obtained by the linear fit of data points is found to be (-1.13388). The electron excitation temperature estimated from the slope of the plot is around 0.88 eV for Argon. The estimation was carried out for three consecutive shots and a small deviation of 0.003 eV is observed. Using theBoltzmann plot method, the electron temperature can be estimated provided the plasma is in LTE and the upper energy level gap of the lines used for the plot should be many times greater than the electron temperature (KT$_e$). This clearly shows that for plasma having high electron temperature, the Boltzmann plot method cannot be used for estimation of electron temperature and gives the electron excitation temperature [37].

## 3.2 Plasma stream evolution

The characteristic of the plasma stream changes from its initiation till its quenching. The evolution study of the plasma beam is important to know any variation in the characteristics of the plasma and its constituent particles from the initiation of plasma formation to the end of its lifetime. The characteristics of such plasma stream at different time of its evolution can be understood with OES method by capturing emission spectra at different times. This is done by triggering the spectrometer to record the spectral data at different times using a delay generator with respect to plasma initiation at breakdown and this allows the emission spectrum to be obtained at different time window. The shots are taken by applying delay times of 50 µs, 100 µs, 150 µs, 200 µs, 250 µs and 300 µs to the triggering of spectrometer from the initiation of the plasma produced by 8 kV discharge pulse. Fig. 5 shows the emission spectrum obtained for different delay times. Since the trigger point of the spectrometer starts the recording of an emission, hence it can be noted here that with 50 µs delay will contain all data of the emission spectrum of the plasma beam excluding the initial duration of 50 µs. Similarly for 100 µs, 150 µs, 200 µs, 250 µs and 300 µs delay also, the emission spectrum will be collected for other time windows of plasma life excluding the initial time frames of 100 µs, 150 µs, 200 µs, 250 µs and 300 µs respectively. This provides information depending on the dominant emissions during that time window. It is observed that with increasing delay, the excitation and upper-level population decreases which indicates the decrease of energy and dying out of plasma. Strong emissions of singly ionized argon ions from different upper excitation levels having both high and low energy value could be observed till 200 µs from the initiation of plasma formation. After 200 µs and onward, transitions occur mostly from low excited upper energy level which is observed in the emission spectrum. This indicates the decrease in electron energy in the system which results in less energy transfer to the ions and atoms by electron collisions. After 300 μs, no prominent emissions are observed as the intensity of the lines is very less and the spectrum appears to be noisy. The spectrum obtained at 350 μs delay appeared noisy and hence cannot not be considered for analysis. Contrary to the change in intensity of ionic lines (Ar II), no significant change in intensity of neutral argon atoms (Ar I) is observed during the entire time window. Transitions of Ar I spectral lines obtained in the emission spectrum are due to radiative decay from $3p^54p$ levels to $3p^54s$ levels. The singly ionized argon (Ar II) line transitions are from $3p^44p$ to $3p^44s$ levels and few are from $3p^44p$ to $3p^43d$ levels. The emissions from the higher excited states are not observed towards later time due to absence of high energy electrons in the plasma. Hence, at this stage electrons does not have sufficient energy to excite the atoms to higher states.

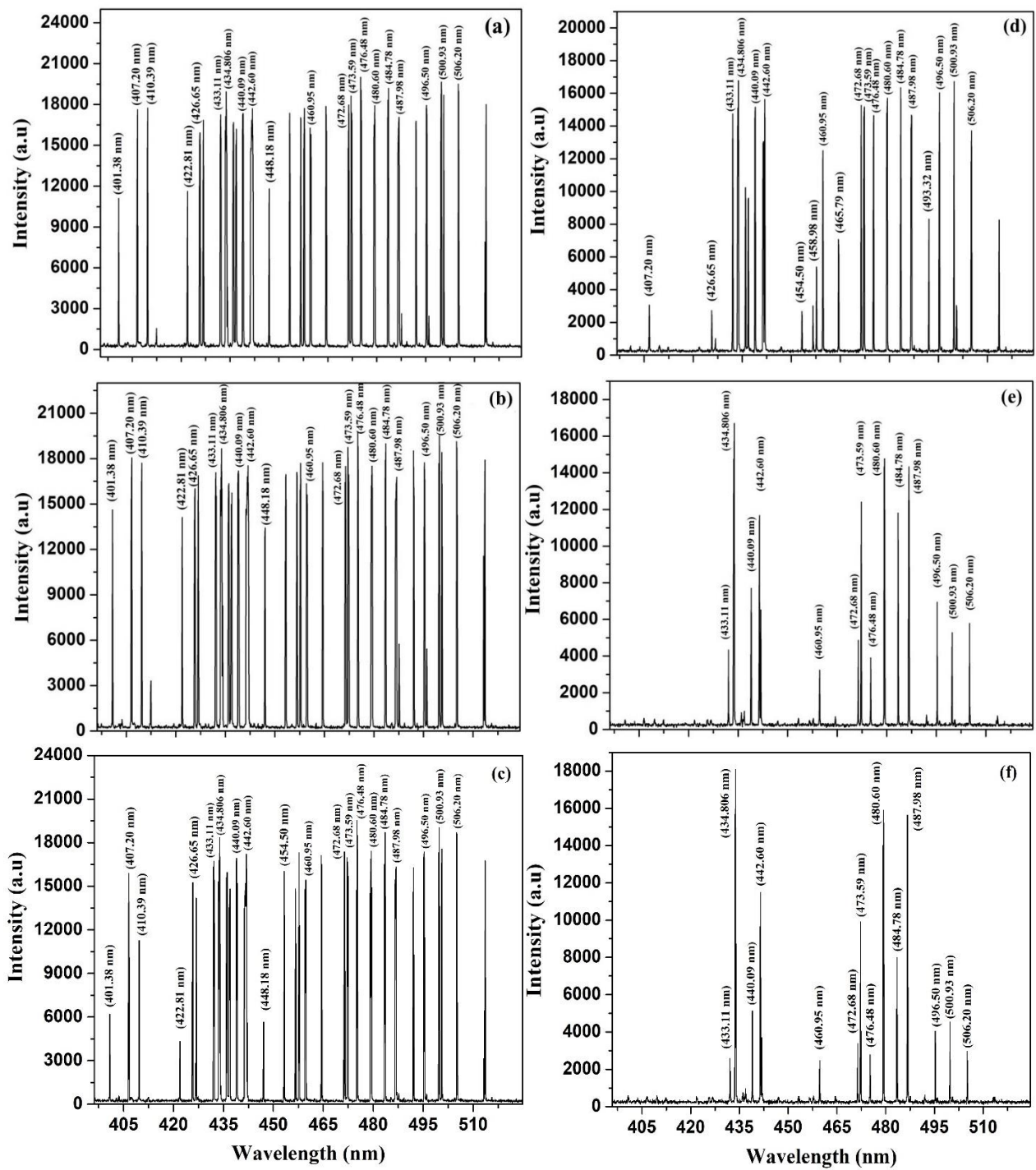

**Fig. 5. Emission spectra of Argon obtained at different delay times of (a) 50 μs (b) 100 μs (c) 150 μs (d) 200 μs (e) 250 μs (f) 300 μs showing the variation of Ar II intensity and evolution of different spectral emissions**

In this study we have also observed the role of metastable states in the emission of plasma. Using selection rules $\Delta L = 0, \pm 1$, $\Delta S = 0$ and $\Delta J = 0, \pm 1$ for optically allowed transitions, the metastable states of singly ionized argon are identified. The emission spectrum reveals the presence of transitions to lower levels which are metastable states having a long lifetime ($10^{-3}$ sec to few seconds) and decay from these states are not radiative. Very few transitions occur to lower excited level which are not metastable states. The allowed transitions from upper excited levels to metastable levels occur by spontaneous emission whereas forbidden transitions occur from the metastable levels to the ground state by collisions. The allowed transitions are radiative and forbidden are non-radiative transitions. Excitations from the metastable states are more probable and involve less energy. But excitations from ground state require much more energy and hence are less probable. Moreover, the electron-impact excitation from the metastable states have greater cross-sections than that from the ground state. Due to this, the intensity of the transitions occurring to metastable lower energy level decreases gradually with time. This is observed in the spectrum where the intensity of transitions occurring to non-metastable levels decreases sharply with time. Fig. 6 shows the intensity variation of few lines making transitions to metastable and non-metastable states, with different delay times. As obtained in the evolution study of the pulsed plasma stream, the argon atoms are singly ionized during the first 100 μs from the initiation of plasma formation.

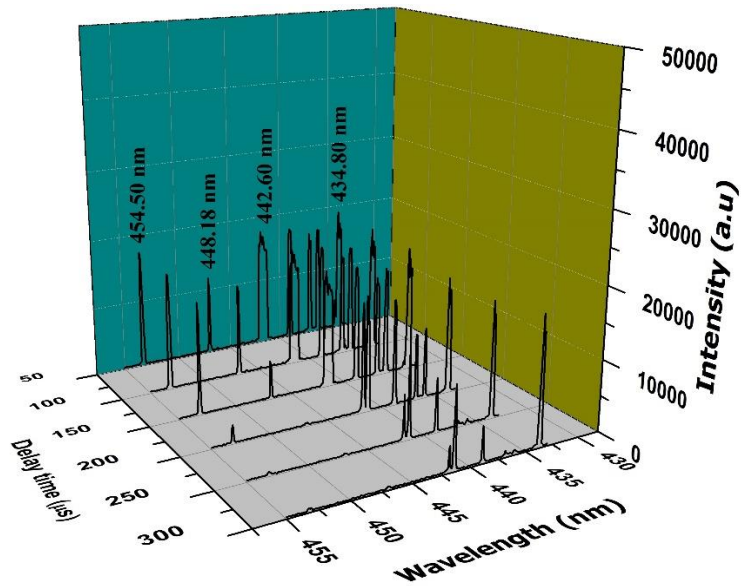

**Fig. 6. Spectral line of Ar II (454.50 nm, 448.18 nm) making transitions to lower non-metastable level whereas Ar II (442.60 nm, 434.80 nm) making transitions to lower metastable level. Figure shows the evolution of the lines with increase in delay time.**

This is confirmed as during that time window, optically allowed transitions to the metastable states could be observed from singly ionized argon atoms. Beyond 100 μs, it is seen that the intensity indicating the population of the argon states in the metastable levels decreases which indicates that no further ionized argon atoms are formed and the ions are only excited from their metastable states. With increase in time delay, depopulation of the metastable states occurs through collisions making non-radiative decays to the ground state.

### 3.2.1 Temporal evolution of Electron Density

Using stark broadening as discussed above, the plasma density is estimated from emission spectra corresponding to different delay times. The spectral line 454.50 nm is considered for density calculation for different successive delay time up to 150 µs. Beyond 150 µs, the intense line 484.78 nm is considered for density calculation, which is free from dip structures at this time window. Using the different electron impact parameter for the corresponding lines selected, the densities are estimated at different times.

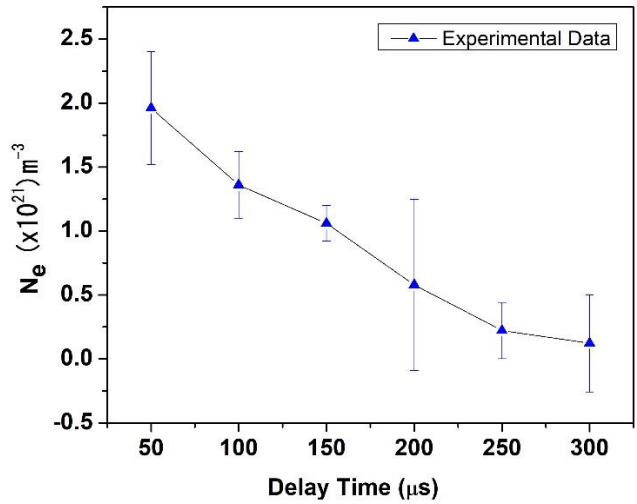

**Fig. 7. Plasma density variation as a function of different delay times**

The accuracy in estimating the density using different spectral lines for different times is checked by calculating the error that arises. This is done by considering two more lines 457.93 nm ad 496.50 nm in addition to 454.50 nm line at a particular time upto 150 µs as plotted in figure 7. After 150 µs, error estimation is done by considering the lines 487.98 nm and 480.60 nm in addition to 484.78 nm spectral line. It is seen that the plasma density decreases with time from $1.96 \times 10^{21}$ m$^{-3}$ estimated at 50 µs to $1.23 \times 10^{20}$ m$^{-3}$ at 300 µs. With the decay in discharge current pulse, the energy input in the system decreases. Moreover, the decrease in current density (due to decaying discharge pulse) results in decrease of Lorentz force acting on the charged particles in the plasma. This reduces the compression of the plasma stream at the exit of the electrode assembly and hence, the plasma density decreases with delay time.

### 3.2.2 Temporal evolution of electron excitation temperature

The change in electron excitation temperature is also determined using the Boltzmann plot method as described above. With increase in time, less number of excited levels follow a Boltzmann distribution as the density decreases and energy transfer occurs only between particles having greater population.

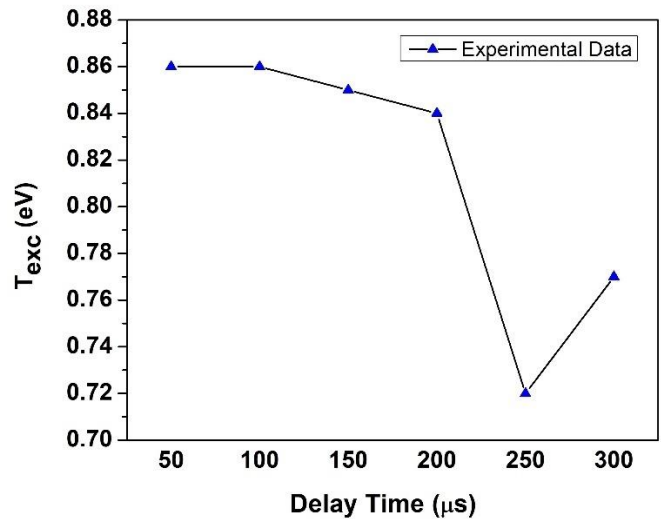

**Fig. 8. Variation of electron excitation temperature with different delay time**

Initially, 7 lines were used for the plot but later some of the excited level populations deviates from the equilibrium state and the excitation temperature is obtained from those levels still in equilibrium. The decrease in excitation temperature at different time window is shown in figure. 8. The decrease of temperature reveals that there is less transfer of electron kinetic energy to the atoms and hence excitation from higher excited levels could not be observed towards later delay time. However, as observed in figure 8, there is an increase in excitation temperature at a delay of 300 μs and this is reflected as increase in intensity of highly populated spectral lines. This is probably due to energy transfer to the electrons during the decrease of plasma density. Similar observation was obtained in an earlier experiment in this device with TLP [29].

4. **Conclusions**

Time-integrated spectroscopic measurements of plasma flow velocity, plasma density, and electron excitation temperature is presented for discharge voltage of 15 kV and 4 bar plenum pressure of gas injection valve for simultaneous triggering of the optical device and the PPS. The plasma flow velocity is found to be $22 \pm 5$ km/s, the plasma density to be $(5.29 \pm 0.38) \times 10^{21}$ m$^{-3}$ and the excitation temperature to be 0.88 eV. This high-velocity plasma stream achieved in this device may be suitable for plasma fueling for fusion devices. The high density of plasma gives a high energy flux which can be utilized for PMI study. The excitation temperature is significant as it reveals the capability of effective kinetic energy transfer of the highly mobile electrons to heavy plasma species (ions and neutrals) through collisions. The evolution of plasma in different time window of its life is studied by adjusting the delay to the trigger of the spectrometer and the information collected from the spectrum shows a variation of plasma density, electron excitation temperature with respect to time. The plasma density has a gradual dcrease from $1.96 \times 10^{21}$ m$^{-3}$ estimated at 50 μs to $1.23 \times 10^{20}$ m$^{-3}$ at 300 μs while electron excitation temperature has shown a rise towards the end of the plasma life. This change is attributed to energy transfer to electrons during decrease of plasma density. This characterization of plasma stream in a pulsed plasma accelerator shows the presence of metastable species which is significant for material processing due to the high reactivity of the metastable species having a long lifetime.

**Acknowledgements:**

The authors are grateful to the Director, Institute for Plasma Research (IPR) and Acting Centre Director, Centre of Plasma Physics- Institute for Plasma Research (CPP-IPR) for supporting us to carry out the present work. The authors also thank Mr.K.K.Kalita and Mr. N.Kathar for assisting in laboratory activities.